\documentclass[Afour,sagev,times]{sagej}

\usepackage{xurl}

\usepackage{moreverb,url}

\usepackage[colorlinks,bookmarksopen,bookmarksnumbered,citecolor=red,urlcolor=red]{hyperref}

\usepackage{multicol}

\usepackage{xurl}

\def\volumeyear{2027}

\usepackage{listings}
\usepackage{xcolor}

\definecolor{codegreen}{rgb}{0,0.6,0}
\definecolor{codegray}{rgb}{0.5,0.5,0.5}
\definecolor{codepurple}{rgb}{0.58,0,0.82}
\definecolor{backcolour}{rgb}{0.95,0.95,0.92}

\lstdefinestyle{mystyle}{
    backgroundcolor=\color{backcolour},
    commentstyle=\color{codegreen},
    keywordstyle=\color{magenta},
    numberstyle=\tiny\color{codegray},
    stringstyle=\color{codepurple},
    basicstyle=\ttfamily\footnotesize,
    breakatwhitespace=false,
    breaklines=true,
    captionpos=b,
    keepspaces=true,
    numbers=left,
    numbersep=5pt,
    showspaces=false,
    showstringspaces=false,
    showtabs=false,
    tabsize=2
}
\begin{document}

\nocite{*}

\runninghead{R D Gill}

\title{Critique of ``Use of roster charts in the investigation
and prosecution of nurses \dots'' by John O' Quigley. }

\author{Richard D. Gill\affilnum{1}
 }

\affiliation{\affilnum{1}Leiden University, Netherlands
}

\corrauth{Richard D. Gill,
Leiden University, Netherlands}
\email{gill@math.leidenuniv.nl
}

\begin{abstract}
The paper ``Use of roster charts in the investigation and prosecution of nurses suspected of inflicting deliberate harm on patients'' by Prof.\ John O'Quigley \citep{oquigley2025} explores an interesting hypothesis concerning statistical information hidden in the part of the infamous Lucy Letby roster chart pertaining to the 37 other nurses. Unfortunately, we have to point out some serious errors in his statistical analyses. The data actually contains information which strongly disproves his main modelling assumption. We do however strongly agree with him that from a forensic statistical point of view, the roster chart is fake evidence which should not have been shown to jurors.
\end{abstract}


\maketitle

\section{Introduction}

Prof. John O’Quigley has published a complex and fascinating paper \citep{oquigley2025} in this journal concerning the nursing roster chart which played a significant role in the controversial case against nurse Lucy Letby. The idea of the paper is that, in fact, such a chart could have been constructed for any other nurse who worked a large number of shifts, if the characterisation of events happening during shifts as suspicious has been biased by suspicion of that particular nurse.

\begin{figure*}[t]
\includegraphics[width=6.75in]{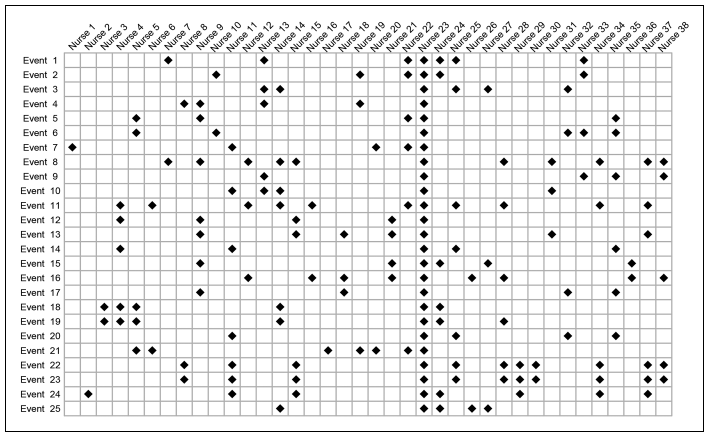}
\caption{A chart believed to show a strong statistical correlation between the presence of nurse 23 and suspicious events. The visual
impression that ignores bias is very strong.}
\label{fig:figure1}
\end{figure*}

He makes this deduction from a statistical analysis of the numbers of events at which each other nurse was present. The idea is that nurses are heterogeneous, some work many more shifts than others, or tend to work different kinds of shifts. We can estimate the degree of heterogeneity from the empirical data on the other nurses. The infamous roster chart has information on 38 nurses, so there are 37 nurses about whom there is no suspicion. They are present at varying numbers of the 25 shifts which contained events in the initial indictment against Lucy Letby.

We are told that originally, hospital investigators “nominated”, as it were, a total of 61 shifts as containing potentially suspicious events. The exact history and exact number are, of course, important, but let us follow this account for the time being. In this paper we are interested in methodological issues (does the method make sense?) and in verifying the computations which have actually been done (given the methodology, are the calculations correct?). Unfortunately, we uncover serious problems, of both kinds.

O'Quigley's idea is that if an event could be considered “unexpected” because the baby it happened to had been considered stable and improving, and no signs of deterioration had been seen, it would be placed on this list of 61. We know that police had asked the hospital to supply medical notes concerning all deaths, and furthermore all unexpected and inexplicable collapses on the unit, in a certain period. They had been told that both numbers were inexplicably large. The selection was presumably made by the same medical consultants who long had suspicions concerning nurse Lucy Letby and had, at a certain point, approached the police.

The records were inspected by retired paediatrician Dr Dewi Evans, who had volunteered himself to the police, via a personal contact at the National Crime Agency, after seeing the police investigation announced in a Sunday newspaper. He identified a large number of, in his opinion, highly suspicious events, and conjectured malicious means whereby the collapse or death could have been caused; for instance, an air embolism caused by deliberate injection of air into the bloodstream. 

Next, the roster data of all 38 nurses was used to build an initial version of the roster chart. A couple of further events were later found by hospital doctors -- the insulin cases. Some revisions were made; the result was 25 events with Lucy Letby present every time. O'Quigley appears to hypothesise that Lucy had not been present at any of the remaining 36 = 61 – 25 events: anyway, for whatever reason, they were no longer considered to be actually suspicious.

The prosecution claimed that Lucy Letby had been present at every event which she was charged with having maliciously caused, and that these events were
all of the actually suspicious events. The initial set of 61 were only ``potentially suspicious''. Dewi Evans had masterfully identified the actually suspicious (because inexplicable) events, and amazingly, just one nurse turned out to have been present, every time.

We don’t know what the roster chart looked like for all 61 candidate events but under some statistical assumptions we can hypothesise what it could have looked like. The 37 ``other nurse'' columns would have varying numbers of X’s in the 36 extra rows (61 = 25 + 36). A statistical analysis of the variation in the number of X’s in the 37 ``other nurse'' columns of the first 25 rows should inform us what the rest of the chart could have looked like. O’Quigley’s statistical analysis purports to prove the following: the variation between the 37 innocent nurses’ predicted total “score” in all 61 shifts is so large that it would be very likely that several other nurses would have scored a total of around 25. Thus, Lucy could merely have been unlucky in being the nurse with the largest number of X’s in the extended, 61-row table. One nurse has to have the most! If “the most”, by chance, could easily be about 25, then another nurse could easily have been selected as “the common denominator” behind the enormous number of deaths and collapses. The roster chart cannot support the hypothesis that a killer nurse is behind the spike in deaths and collapses, let alone that it is Lucy Letby.

In principle, that might have been the case. However we will now identify flaws in the published analysis – both in the model assumptions and in the execution of computational statistical analyses. This does not mean that the work was done for nothing. Much can be learnt from it, and this directs further analyses which others are doing, and which further do completely support O’Quigley’s main message: the roster chart is deeply misleading. 

In our opinion, it is true that Lucy Letby was unlucky. But her bad luck had a systematic cause: she was systematically rostered much more often in shifts where serious adverse events were likely to happen, through standard protocols regarding the necessary composition of nursing staff present in any shift, and through her own preparedness and enthusiasm to do shifts when the unit otherwise would be short-staffed. The problem is that the composition of the nursing needs of patients on the unit is not independent of the composition of nurses’ qualifications needed to provide adequate care for those patients. Nurses are indeed heterogeneous. But so are shifts. Shifts and nurses are not independent. But O'Quigley's main assumption is that they are completely independent.

Could it still be approximately true, so that his analysis is still useful? As we will see, it is very far indeed from being true, and, had his statistical calculations been correct, he would have noticed this for himself.

\begin{figure*}[t]
\includegraphics[width=6.75in]{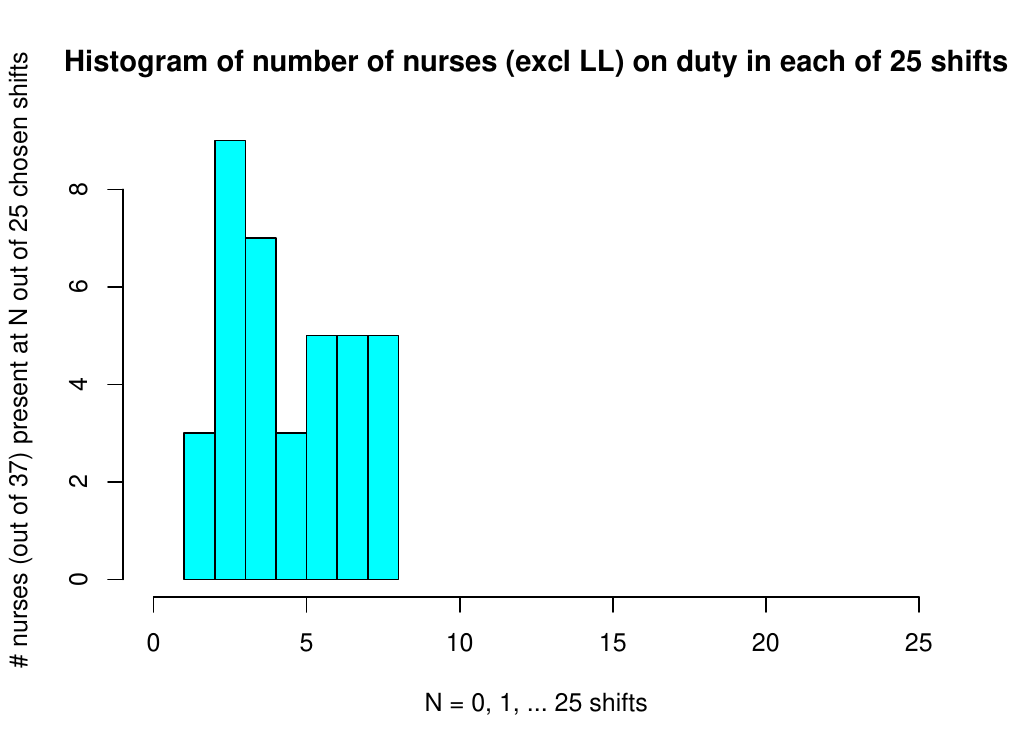}
\caption{The histogram of the raw data for the presence of the unsuspected nurses.}
\label{fig:figure3}
\end{figure*}

\section{O'Quigley's assumptions}

The starting point is the familiar roster chart, as reproduced by \citep{oquigley2025} as his ``Figure 1''. It uses a diamond instead of an ``X'' to mark a nurse's presence in a given shift.  We have duplicated it (with his caption) in our Figure \ref{fig:figure1}. He explains that he is going to model the data in the 37 columns remaining when we exclude Lucy Letby's column (column 23). The columns are ordered by alphabetical order of surname. The order of the rows is almost chronological, and we can easily see the effects of time dependence, as well as another characteristic of the shifts: day shift, or night shift. During the daytime (and especially on weekdays), a number of nurses are doing administrative duties. Day or night, weekend or weekday, roughly the same number of nurses are actually allocated personally to the care of one or two specific babies, and are supposed to be continuously responsible for them.

There are lots of what look like small vertical clusters of a few X's in neighbouring rows and columns. The same nurse is often present when events happen to one child, close together in time. Twins and multiples tend to have events happening very close by in time and the same nurses are often there. Their columns are adjacent to one another.

O'Quigley has an audacious solution to the just mentioned problems: randomise the order of the rows (incidents, babies), \emph{forgetting} all information about date and time; thereby also discarding information of several babies having the same mother, and multiple events happening to the same baby. Information about correlation between events is made inaccessible. His idea is that the hospital has identified a total of 61 potentially suspicious events. Through a process of re-evaluation of medical events, 25 have been determined to be actually suspicious, and Lucy Letby is seen as the common factor in these 25. He believes that the process was biased towards seeing events as suspicious if Lucy were present. The prior selection of unexpected and unexplained, but not immediately fatal, collapses was performed by the NNU consultants and therefore was very likely biased against Lucy Letby.

\begin{figure*}[t]
\includegraphics[width=6.75in]{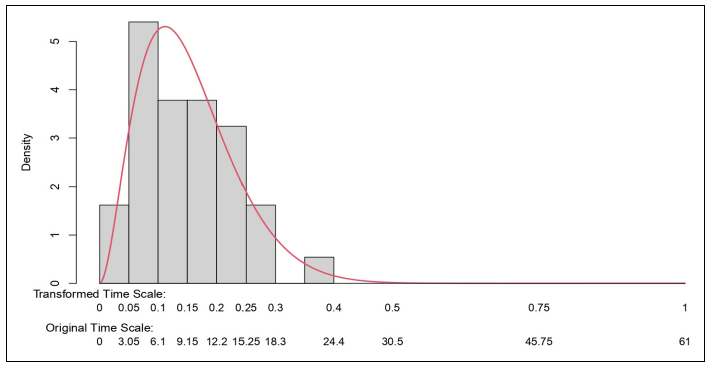}
\caption{O'Quigley's Figure 4. His caption: A histogram showing the raw data for the presence of the unsuspected nurses [{\bf Not true!} {\it This is not a histogram of the raw data}]. A beta-binomial distribution is fit to the data using maximum likelihood estimates [{\bf Not true!} {\it Maximum likelihood estimation converges to the degenerate case: the binomial}].}
\label{fig:figure2}
\end{figure*}

Now we know that there are many different kinds of nurses and some actually work many less hours per week than Lucy. Perhaps only a handful of nurses work as many hours as she does. He hopes to show that any of this group could easily also be present at around 25 of the 61 candidate suspicious events. Thus if doctors had early on become suspicious of a different nurse, they might just as well have ended up with an essentially identical roster chart, dramatically putting the blame on that other nurse.

O'Quigley's fundamental assumption (assuming Lucy's innocence) is that the X's in the original 61 row spreadsheet, \emph{after} randomly shuffling the order of the rows, are all placed at random, independently of one another, but using a different probability of being rostered for each of the 38 nurses. The reshuffling is a controversial move. It finesses the problem of the mixture of day and night shifts, and the problem of the time correlation of events happening to one child (or to siblings). 

We do not know nurses' individual probabilities of being present in the 61 candidate events. O'Quigley seems to have two different ideas concerning this. One is that each innocent nurse's chance of presence at each of the $36 = 61 - 25$ not-selected candidate events may be taken to be equal to their observed relative frequency of presence in the $25$ selected events. He seems to think this is justified by statistical principles for dealing with missing data. This author completely disagrees. The justification which O'Quigley finds in Rubin and Little's concept ``missing completely at random'' (MCAR) has no mathematical basis at all. We will come back to this topic when we get to the very last part of O'Quigley's paper where he turns to what might be called a non-parametric approach. In a recent talk he further develops this line of attack, now associating it with Rubin's Bayesian bootstrap methodology \citep{rubin1981}; in our opinion, equally spurious, unjustifable.

He first explores another idea which seems to me, a priori, certainly worth trying: the beta-binomial model. This could be called a parametric approach. Each nurse has their own ``long term'' relative frequency $p$ of being rostered in a shift.  These 38 unknown numbers are modelled as a random sample of size 38 from a flexible two-parameter family of probability distributions called the beta distribution.
We can estimate the two parameters of that distribution from the sample of 37 innocent-nurse presence-totals in 25 shifts -- the 37 column sums of Figure \ref{fig:figure1}, excluding Lucy's column. In our Figure \ref{fig:figure3}  is the actual histogram of that actual data.

We can now simulate the data-generating mechanism behind Figure \ref{fig:figure1} as follows (assuming innocence of all nurses): create a sample of size 38 $p$'s from the beta distribution estimated previously; sample an X or a blank for the 61 shifts for each nurse independently, using that nurse's simulated $p$. Now look for nurses with a large column total. O'Quigley's prediction is that there will be two or three nurses with around 25 presences. Pick one of them, and decide that all of those events are truly suspicious. Delete the other rows, and we have a spreadsheet like the one used to convict Lucy Letby.

\begin{figure*}[t]
\includegraphics[width=6.75in]{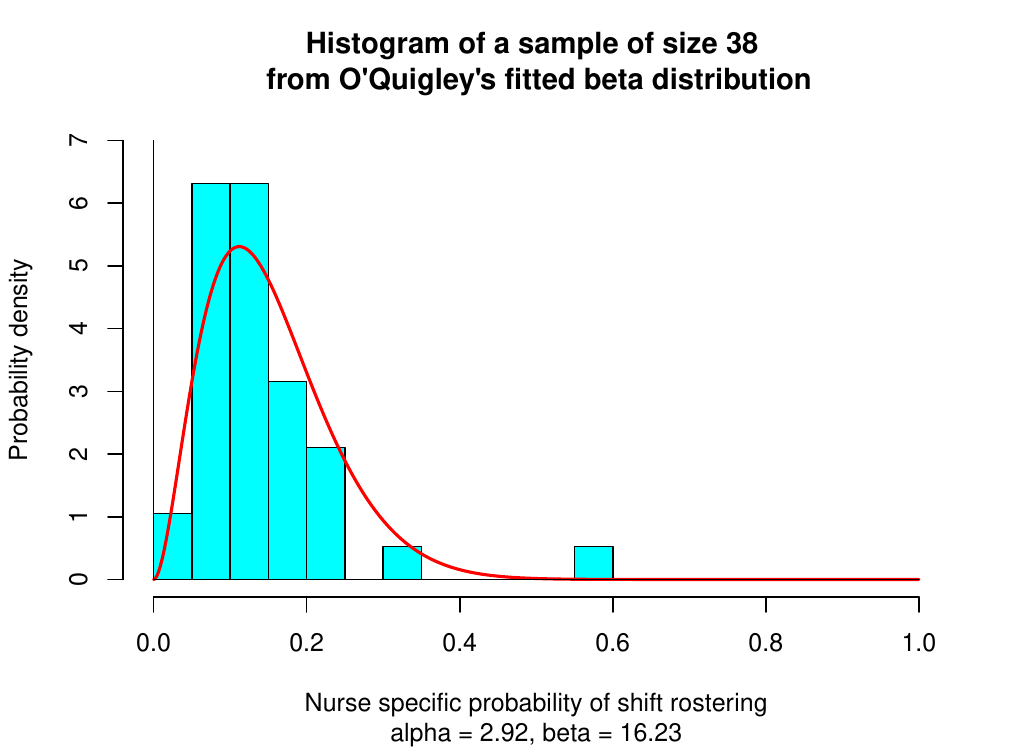}
\caption{Histogram of a sample of size 38
from O'Quigley's fitted beta distribution, with the estimated density
superimposed. The total area of the bars equals the total
area under the curve.}
\label{fig:figure4}
\end{figure*}

O'Quigley states that his Figure 4, which is reproduced here as our Figure \ref{fig:figure2}, ``shows Condition 3 [the beta-binomial assumption] to provide a good fit to the observations.
In fact, it does no such thing! The histogram (bar plot) is not what he says it is. It is definitely not a histogram of the sample of the 37 ``other nurse'' column totals in the 25-row chart given to the press. For that, see our Figure \ref{fig:figure3}.  It is something else entirely. I suspect the data summarised in the histogram comes from a simulation of a sample of size 37 or 38 from the estimated beta distribution, which would justify plotting the continuous beta density and the normalized histogram (normalised so that the total area of the bars equals one, just like the area under the density curve). Every time one takes a new sample from the same beta density, one will get a different histogram. For another one, see our Figure \ref{fig:figure4}.

O'Quigley calls the horizontal axis a time axis, apparently, seeing 61 events as occurring spread out in time. But the figure supposedly is about numbers of nurses having different numbers of shifts. This makes no sense at all.

We can at least identify \emph{which} beta distribution is used here. There exists an archived pre-publication version, \citep{oquigley2024}, of what finally became the \citep{oquigley2025} paper. It has just a few more details and allows us to  identify the beta distribution which he had found -- more precisely, which his computer-savvy colleague biostatistician Dr.\ Sean Devlin had found. Good statistical computing practice is
to save R scripts (programs written in the language R) 
and to have them include the random seed, set in advance, 
so that Monte Carlo simulations can be \emph{exactly} reproduced (a different matter from simply replicating a Monte Carlo simulation by generating new, independent random samples). By definition, Monte Carlo calculations have sampling error. It seems that Devlin had sent O'Quigley R scripts written for the project, but O'Quigley seems not to have kept them.

This brings us to the next anomaly.
When we tried to fit the beta binomial distribution ourself, following O'Quigley's fairly explicit instructions, something very odd happened: the maximum likelihood fitting algorithm failed to find a solution. The likelihood was apparently maximised at the boundary of the parameter space, where $\alpha$ and $\beta$ both converge to infinity in such a way that the smooth beta mixture of binomials converges to the degenerate case: all mass on just one value of $p$. A single binomial fits the data better than a beta mixture of binomials.

The method of moments did produce a proper beta distribution solution. However, $\alpha$ and $\beta$ were both so large that this solution too was effectively the plain degenerate case preferred by the maximum likelihood method. The method of moments fit was possible because the actual sample variance, though smaller than the actual sample mean, was larger than $25 * p * (1 - p)$, where $p$ is the solution of the estimating equation ``sample mean $ = n p$'' with $n = 25$.

In our Figure \ref{fig:figure4},  is a graph of O'Quigley's fitted beta density together
with a histogram of a random sample of size 38 from that 
probability density. In that run, just one nurse
got a $p$ of nearly 0.6, hence they would likely be present for
at least 30 of 61 shifts, and their excessive presence would stick out
like a sore thumb. In other runs, the result was more often that there were two or three nurses with $p$ around 0.4, and therefore several nurses with around 25 shifts. Had the data actually fit well to this model, O'Quigley might have nicely achieved all his aims.

But as said, the beta distribution used in O'Quigley's paper is \emph{not} the  maximum
likelihood fit of a beta-binomial distribution. We hypothesise the following: O'Quigley has sampled the 36 missing rows of the table by using the 37 other nurses's observed relative frequency of presence in the columns of the 25 observed rows, and done maximum likelihood
estimation of the beta-binomial using the augmented table of 61 rows. This seems to match to his
idea of using Little and Rubin's MCAR hypothesis (or principle). Each new Monte Carlo simulation will generate different estimates. Did Devlin average the results of many 
replications? O'Quigley doesn't say and possibly does not even know. Devlin discarded old computer files when he changed jobs a little while ago. 
He does not remember fine details of the collaboration. 

Our hypothesis does explain the overdispersion predicted by the estimated model. The observed relative frequencies for each nurse  are only estimates of each nurse's  ``true'' long term rate of presence. Using the estimates  instead of the true values increases the variance of the set of 37 or 38 probabilities and hence increases the variance of nurse column totals. The average is not affected. In the original data, average is approximately equal to sample variance.  But we now end up with a fitted model with variance about twice as large as the mean. In several experiments we found very similar beta parameters. 

However it was obtained, this particular estimated beta distribution generates
a rather large degree of heterogeneity between the nurses. The resulting
probability distribution of 37 random nurses' random column totals
in a $25 \times 37$ roster chart would exhibits strong overdispersion: variance much larger than the mean. The actual sample variance of the 
37 innocent nurse's 25 shift presence totals in the actual roster chart  is smaller
than the sample mean. The roster chart column totals (25 shifts, 37 innocent nurses) exhibit \emph{underdispersion}, not \emph{overdispersion}!

The thing is, that the number of nurses actually caring directly for babies in incubators is pretty small and directly linked to the numbers of babies in the different sections of the NICU. In busy periods, once you have rostered nurses
of the necessary qualifications and in principle available for a shift, other nurses are not needed for intensive patient care. Thinking of nurses as atoms in a physical model of some kind of matter, where different atoms are located at different locations, one would say that ``nurses repel one another''. You can't have very many very similar nurses on the same shift.

At this point, possibly enough has been said. The model used by O'Quigley has been proven
to be seriously wrong, and the paper's description of the findings is inadequate. 
Figure captions and figure descriptions are completely wrong. There has been a huge mixup. However, we will continue to the bitter end.

In the final part of O'Quigley's paper, the author claims to show that Lucy Letby's number of suspicious events is actually typical for the highest scores among nurses working as many hours as she does. This is a true fact for his estimated beta-binomial model. A random nurse has a probability of 1.3\% of having a total of 25 presences in 61 rows (see O'Quigley's Figure 5). It follows that the probability that all 38 nurses would have less than 25 is about 60\%. 
It follows that, as likely as not, the highest scoring nurse will have around 25 presences.

O'Quigley prefers a complicated and, to our mind, superfluous argument. He is an expert in survival analysis and 
prefers to segue to the proportional hazards model. His shuffled 61 events are now thought of as 61 events uniformly distributed in time.
He writes ``nurse 23 [Lucy Letby] is a member of the group of nurses whose time on duty equals or exceeds by more
than 33\% the overall average ... she actually works 35\% more hours than the average nurse.''
He does not give precise specification of what exactly he has plotted in the four panels of his last and very important Figure 6. 
The present author considers himself also as an expert in survival analysis, but quite simply has no idea at all what the
exact specifications of the curves in O'Quigley's Figure 6 are. A recent conference presentation by O'Quigley, published in a 
YouTube video, presents them together with some new figures based on new analyses
using the Bayesian bootstrap methodology of \citep{rubin1981}. He says these will be
presented in an upcoming paper entitled ``The elephant in the room''.

\section{Conclusions}

\begin{figure}[ht]
\includegraphics[width=3in]{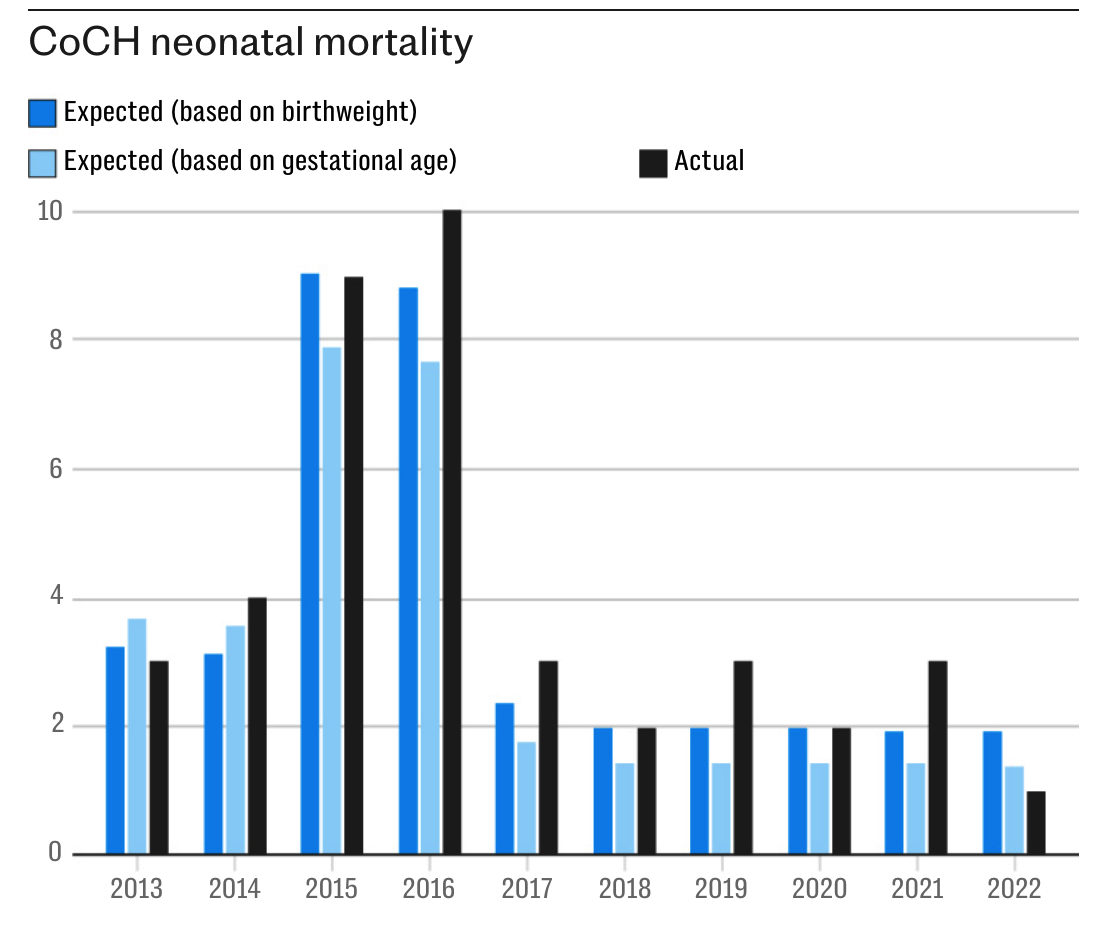}
\caption{The expected deaths series are derived by combining three data series: a) generic mortality risk by birthweight and gestational age b) CoCH births distribution by birthweight/gestational age assuming in line with general distribution (adjusted in 2013-14, 2015-16 and 2017-22) c) CoCH actual number of births. It is designed to show the high sensitivity of mortality to small shifts in birthweight/gestational age and how the actual mortality can be reproduced closely. Reproduced from \citep{knapton2024}.}
\label{fig:figure5}
\end{figure}

Quite an enormous amount went wrong in \citep{oquigley2025}; and a lot can be learnt from that. As a paper written for a forensic science journal of which not many readers or editors would be persons at the intersection of mathematical
statistics and applied statistics, the author used a breezy style in which
he ran over technical matters quite rapidly. His main message, that the
roster chart is a scientific fake, came across clearly. We completely agree that that is the case. We already know that from the actual history of how the
chart was constructed and from the work of Peter Elston and others, reported in YouTube videos and newspaper articles, cf.\ \citep{knapton2024} and Figure \ref{fig:figure5}. O'Quigley hoped to show that more is true, namely,
that Lucy Letby was just unlucky in the sense that very easily, another nurse could have appeared just as well to stand out in the roster chart. We disagree
that this is the case, and hope to publish on the matter in the near future. But anyway, since the fitted model used by O'Quigley completely fails to 
fit to the actual data, predictions (more precisely, retrodictions) made from that model fit are useless.

Confirmation bias and poor communication between O'Quigley and his programmer,
his former colleague Dr.\ Sean Devlin, seems to have led to a catastrophe. 
It was alas not noticed by editors and referees.
The author(s) should have provided supplementary material containing
explicit mathematical derivations and R scripts.

\section{Note}

The R scripts written by the author for calculations and
graphics in this paper are all available from him on request.
The code which produced Figure \ref{fig:figure4} is reproduced 
in the Appendix below.

\raggedright
\bibliographystyle{SageV}

\section{Appendix}

\begin{lstlisting}[language=R, caption=R script for Figure 4]
Alpha <- 2.92 # From O'Quigley(2024)#
Beta <- 16.23 # From O'Quigley(2024)
library(MASS)
set.seed(123456)
p <- (0:1000)/1000
rbetas <- rbeta(38, Alpha, Beta)
truehist(rbetas, h = 0.05, xlim = c(0, 1), ylim = c(0, 7), prob = TRUE,
main = "Histogram of a sample of size 38 \n from O'Quigley's fitted beta distribution",
sub = "alpha = 2.92, beta = 16.23",
xlab = "Nurse specific probability of shift rostering", ylab = "Probability density")
lines(p, dbeta(p, Alpha, Beta), col = "red", lwd = 2)
lines(x = c(0, 1), y = c(0, 0)); lines(x = c(0, 0), y = c(0, 7))
\end{lstlisting}

The first two lines define variables \texttt{Alpha} and  \texttt{Beta} which are assigned the values 
reported by O'Quigley. The third line loads a standard library called \texttt{MASS}, standing for
\emph{Modern Applied Statistics with S-PLUS}, a text book by B. D. Ripley (1999) which explained the
use of a fore-runner of the present programming language \texttt{R}. I prefer to use his improvement,
called \texttt{truehist}, of R's standard and built in function for plotting a histogram.

The next line sets a random seed. If one simply runs the same code again, exactly the same
random sample will be taken. By replacing the seed with any other number one will get a new sample. 
Re-running the same code several times with a new seed every time will give the user a nice real time impression
of statistical sampling variation. 

The next line defines a vector called \texttt{p} of length 1001 containing equally spaced numbers from 0 to 1 in
steps of size 1/1000. This will be used as a plotting variable – we are going to plot the beta
probability density $y = p(x; \alpha, \beta)$ against its argument $x$, which confusingly is itself a probability.

At last, in the next line, we take a sample of size 38 from this beta distribution. The 38 sample values are contained
in a vector called \texttt{rbetas}.

Now we can plot a histogram of those 38 values. The option \texttt{prob = TRUE} normalizes the histogram so that its bars fill a total area of size 1, and hence can be thought of as a piece-wise constant estimate of a probability density. In the next line of
code we superimpose on the plot the actual beta density from which the values were sampled.
A final line of code adds some boundary lines to the plot. \texttt{R}'s function \texttt{dbeta} computes the beta density, its function \texttt{rbeta}
samples from that distribution,

\end{document}